\documentclass[preprint2]{aastex631}

\usepackage{rotating}

\usepackage{booktabs}
\definecolor{bettergreen}{rgb}{0.0, 0.5, 0.0}

\graphicspath{{./}}

\received{May 23, 2023}
\revised{August 1, 2023}
\accepted{August 14, 2023}
\submitjournal{ApJ}

\shorttitle{Methanol Dasar in The Brick}
\shortauthors{Bulatek et al.}

\begin{document}

\title{The 107~GHz methanol transition is a dasar in G0.253+0.016}

\correspondingauthor{Alyssa Bulatek}
\email{abulatek@ufl.edu}

\author[0000-0002-4407-885X]{Alyssa Bulatek}
\affil{Department of Astronomy, University of Florida, P.O. Box 112055, Gainesville, FL 32611, USA}

\author[0000-0001-6431-9633]{Adam Ginsburg}
\affil{Department of Astronomy, University of Florida, P.O. Box 112055, Gainesville, FL 32611, USA}

\author[0000-0003-2511-2060]{Jeremy Darling}
\affil{Center for Astrophysics and Space Astronomy, Department of Astrophysical and Planetary Sciences, University of Colorado, 389 UCB, Boulder, CO 80309-0389, USA}

\author{Christian Henkel}
\affil{Max-Planck-Institut für Radioastronomie, Auf dem Hügel 69, 53121 Bonn, Germany}
\affil{Astron. Dept., Faculty of Science, King Abdulaziz University, P.O. Box 80203, Jeddah 21589, Saudi Arabia}
\affil{Xinjiang Astronomical Observatory, Chinese Academy of Sciences, 830011 Urumqi, PR China}

\author{Karl M. Menten}
\affil{Max-Planck-Institut für Radioastronomie, Auf dem Hügel 69, 53121 Bonn, Germany}

\begin{abstract}
We present observations of population anti-inversion in the $3_1 - 4_0\ A^+$ transition of CH$_3$OH (methanol) at 107.013831 GHz toward the Galactic Center cloud G0.253+0.016 (``The Brick''). Anti-inversion of molecular level populations can result in absorption lines against the cosmic microwave background (CMB) in a phenomenon known as a ``dasar.'' We model the physical conditions under which the 107~GHz methanol transition dases and determine that dasing occurs at densities below $10^6$ cm$^{-3}$ and column densities between $10^{13}$ and $10^{16}$ cm$^{-2}$. We also find that for this transition, dasing does not strongly depend on the gas kinetic temperature. We evaluate the potential of this tool for future deep galaxy surveys. We note that other works have already reported absorption in this transition (e.g., in NGC 253), but we provide the first definitive evidence that it is absorption against the CMB rather than against undetected continuum sources.
\end{abstract}

\keywords{\href{http://astrothesaurus.org/uat/103}{astrophysical masers}, \href{http://astrothesaurus.org/uat/565}{Galactic Center}, \href{http://astrothesaurus.org/uat/1072}{molecular clouds}, radio lines: ISM}

\section{Introduction}

Microwave amplification by stimulated emission of radiation, called a ``maser,'' occurs when two energy levels in a population of particles are inverted or contain a non-thermal distribution of particles. Particles are ``pumped'' into the upper state until it is over-populated, driving a population inversion. The level populations are summarized by the excitation temperature using the following relation.
\begin{equation}
    \frac{N_u g_l}{N_l g_u} = \exp\left(\frac{-\Delta E}{k_B T_{ex}}\right)
\end{equation}
Here, $N_u$ and $N_l$ are the number of particles in the upper and lower energy states respectively, and $g_u$ and $g_l$ are the statistical weights of those states respectively. $\Delta E$ is the difference in energy between the two states, $k_B$ is the Boltzmann constant, and $T_{ex}$ is the excitation temperature of the transition (for maser transitions, $T_{ex}$ is negative). The inverse effect, where the lower state is over-populated resulting in an anti-inversion, can also occur.

Molecular line absorption against the cosmic microwave background (CMB), which can only happen when the level populations are anti-inverted, has been observed \citep[e.g.,][]{palmer1969formaldehyde}. The term ``dasar'' was coined to describe this phenomenon \citep{townes1997astronomical}. ``Dasar'' stands for ``darkness amplification by stimulated absorption of radiation.'' Dasars, like masers, require a ``pump'' to overpopulate an energy state over the thermal distribution. Unlike masers, they are not enhanced by geometric amplification---the ``amplification'' in the acronym is therefore a misnomer kept only for linguistic symmetry.

A pumping mechanism for dasing was described by \citet{townes1969pumping}. They suggested that the most likely mechanism for overpopulating the lower state of a dasar transition is collisions between heavy spherical atoms (e.g., neon) with the dasing molecule (in this case, H$_2$CO). If the spherical atom strikes the hydrogen atoms in the H$_2$CO molecule in the plane of the molecule, the angular momentum of the molecule increases much more than if the atom strikes the carbon or oxygen atoms, or if it strikes the hydrogen atoms perpendicular to the plane of the molecule. This large increase in angular momentum tends to excite the molecule from the $1_{11}$ and $1_{10}$ states into the $2_{12}$ and $3_{13}$ states instead of the $2_{11}$ and $3_{12}$ states. The $2_{12}$ and $3_{13}$ states prefer to spontaneously radiatively de-excite into the $1_{11}$ state, the lower state of the $1_{10} - 1_{11}$ \textit{K}-doublet. This causes an overpopulation of the lower state. 
\citet{garrison1975cooling} verified this collisional pumping mechanism for H$_2$CO by performing quantum mechanical calculations with helium as the collisional partner.

Several molecules (e.g. H$_2$CO, c-C$_3$H$_2$, c-C$_3$H, c-C$_3$D, and CH$_3$OH) are predicted and observed to dase. H$_2$CO (formaldehyde) absorption against the CMB was first seen in the $1_{11} - 1_{10}$ transition at 4.83 GHz by \citet{palmer1969formaldehyde} in a Galactic infrared dark cloud. \citet{evans1975interstellar} detected the $2_{12} - 2_{11}$ transition of formaldehyde at 14.5 GHz in absorption against the CMB in several Galactic dark clouds. In an extragalactic setting, \citet{mangum2008formaldehyde} presented a survey of the aforementioned formaldehyde dasar transitions in a sample of starbursting galaxies. The $1_{11} - 1_{10}$ formaldehyde dasar was also detected in the extragalactic gravitational lens B0218+357 at $z=0.68$ \citep{zeiger2010formaldehyde}. The $2_{12} - 2_{11}$ and $3_{13} - 3_{12}$ lines of H$_2$CO were detected in PKS1830-211 at $z=0.89$ \citep{menten1999interferometric, schulz2015inhomogeneous}.

Four transitions of c-C$_3$H$_2$ (cyclopropenylidene) have been shown or predicted to appear in absorption against the CMB. \citet{matthews1986transition} detected the 21.6 GHz $2_{20} - 2_{11}$ transition of c-C$_3$H$_2$ in absorption against the CMB in a variety of Galactic sources, including TMC-1, L134N, Sgr B2, and W51. Though they have not yet been astronomically observed, three other transitions of c-C$_3$H$_2$ are predicted to dase \citep[$4_{40} - 4_{31}$ at 35.4 GHz, $3_{30} - 3_{21}$ at 27.1 GHz, and $4_{32} - 5_{05}$ at 19.1 GHz;][]{sharma2022anomalous}.

The radical c-C$_3$H and its deuterated counterpart c-C$_3$D have several transitions which are predicted to dase \citep[for c-C$_3$H, $3_{31} - 3_{30}$ at 3.4 GHz and $1_{10} - 1_{11}$ at 14.8 GHz, and for c-C$_3$D, $3_{31} - 3_{30}$ at 0.9 GHz and $1_{10} - 1_{11}$ at 10.8 GHz;][]{chandra2007anomalous}.

Several transitions of CH$_3$OH (methanol) have been observed in absorption against the CMB. The $2_0 - 3_{-1}\ E$ transition at 12.1 GHz was seen in absorption against two local dark clouds, TMC-1 and L183 \citep{walmsley1988antiinversion}. Additionally, the $5_1 - 6_0\ A^+$ transition at 6.7 GHz was detected in absorption against the CMB in two hot corinos in NGC 1333 \citep{pandian2008detection}.

\begin{figure*}
    \centering
    \includegraphics[width=0.55\textwidth]{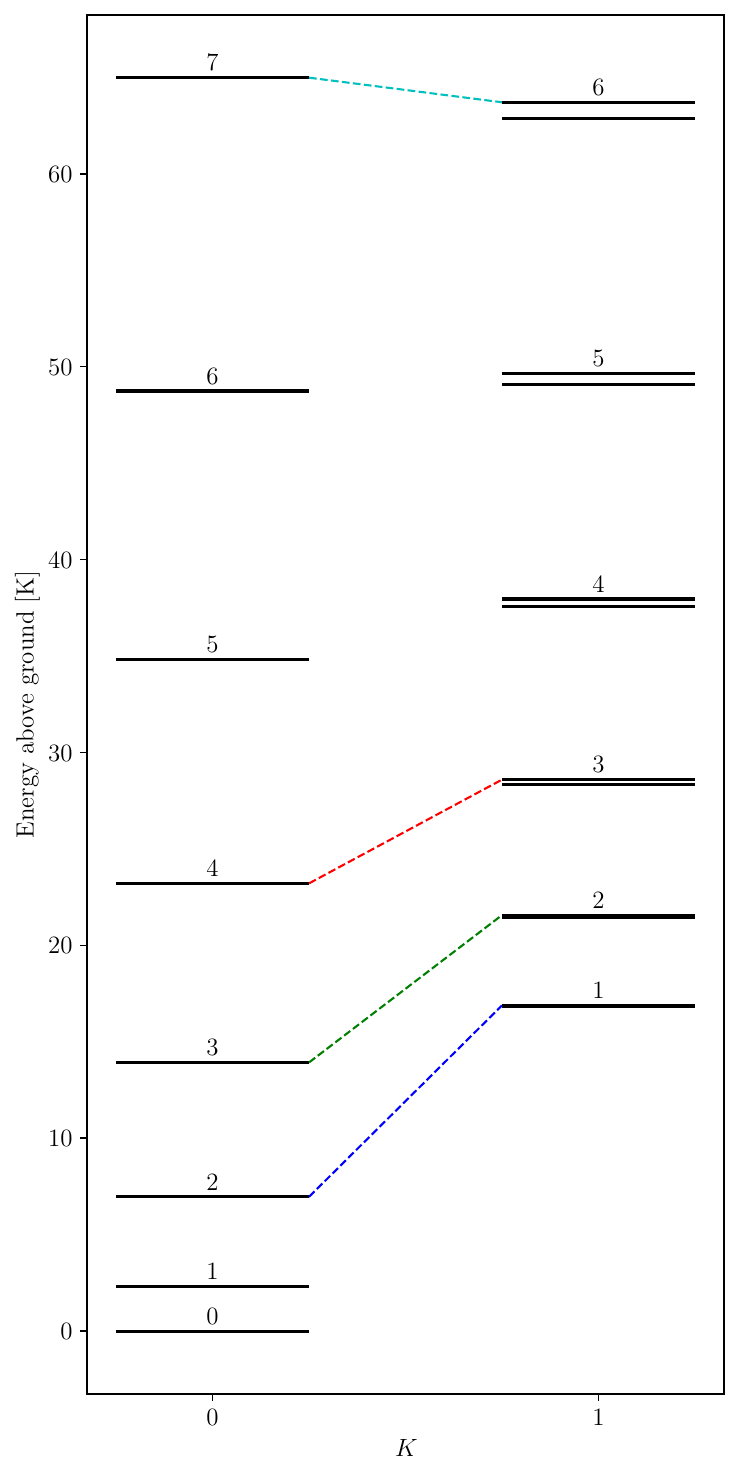}
    \caption{Rotational energy level diagram for the $A$ ladder of methanol up to $\sim$65 K. The energy of a level is represented on the vertical axis, and the horizontal axis represents the $K$ quantum number of the level. Parity splitting occurs for the $K = 1$ levels; the split levels with the slightly higher energies are involved in $A^+$ transitions while the levels with the slightly lower energies are involved in $A^-$ transitions. The numbers above each line represent the $J$ rotational quantum number of the level. The \textcolor{cyan}{cyan} dashed line represents the $7_0 - 6_1~A^+$ transition at $\sim$44 GHz, the \textcolor{red}{red} dashed line the $3_1 - 4_0~A^+$ transition at $\sim$107~GHz, the \textcolor{bettergreen}{green} dashed line the $2_1 - 3_0~A^+$ transition at $\sim$157~GHz, and the \textcolor{blue}{blue} dashed line the $1_1 - 2_0~A^+$ transition at $\sim$206~GHz. Adapted from Figure 3 in \protect\citet{leurini2016physical}.}
    \label{fig:energydiagram}
\end{figure*}

The $3_1 - 4_0\ A^+$ transition of methanol at 107.013831 GHz has previously been reported as a maser, but never as a dasar \citep{valtts1995discovery, turner1991molecular}.
The properties of the 107.013831 GHz methanol transition (henceforth referenced as the 107~GHz or $\sim$107~GHz transition) are shown in Table \ref{tbl:lineinfo}. In Figure \ref{fig:energydiagram}, we show an energy level diagram that highlights the 107~GHz transition.

\begin{table}
\centering
\caption{Information about several possible methanol dasar transitions from CDMS \citep{endres2016, muller2005cologne}. The 107~GHz methanol transition is the main topic of this work. Given quantum numbers are the total angular momentum $J$; indices ($K$) denote the projection of $J$  onto the molecular rotation axis. The $A$ indicates that the three hydrogen atoms connected to the carbon atom have parallel spins.
$A^+$ indicates that the transition is from $A$-type methanol and the $K = 1$ state has a slightly higher energy than the $A^-$ variant.}
\begin{tabular}{ccc}
\toprule
 Rest frequency & Transition & $E_U$ \\
\midrule
       107.013831 GHz & $3_{1} - 4_{0}$ $A^+$ & 28.34820 K \\
       156.602395 GHz & $2_{1} - 3_{0}$ $A^+$ & 21.44383 K \\
       205.791270 GHz & $1_{1} - 2_{0}$ $A^+$ & 16.84073 K \\
\bottomrule
\label{tbl:lineinfo}
\end{tabular}
\end{table}

Dasars may be unique probes of high-redshift galaxies. \citet{darling2012formaldehyde} demonstrated that H$_2$CO dasar transitions are detectable independent of redshift. This is because absorption lines are detectable no matter what distance the absorbing material is from the observer.

The detection of the 107~GHz methanol line as a dasar would provide an opportunity for a methanol dasar ``deep field'' survey that can complement other proposed surveys of dasars in high-redshift galaxies at lower frequencies \citep[e.g.,][]{darling2018formaldehyde}. The 107~GHz line, as well as a few other methanol lines that are expected to dase (at $\sim$157 and $\sim$206~GHz; see Table \ref{tbl:lineinfo}), are located at very convenient frequencies for the next generation of radio interferometers (e.g., the next-generation Very Large Array), especially at higher redshifts. Such a survey would probe regions of star formation in high-redshift galaxies with greater sensitivity than ever before.

In this work, we report the detection of absorption of the CMB by methanol at 107~GHz in the Galactic Center molecular cloud G0.253+0.016, also called ``The Brick.'' In Section \ref{sec:data}, we describe data acquisition and reduction. In Section \ref{sec:detection}, we demonstrate that the 107~GHz transition is a dasar using spectra extracted from areas in and around the region with dasar absorption. In Section \ref{sec:radiative}, we explore radiative transfer modeling to estimate the physical conditions that could enable dasing. In Section \ref{sec:applications}, we discuss the cosmological applications of this new dasar as well as its detection in a nearby starburst galaxy. In Section \ref{sec:conclusions}, we summarize the implications of our results.

\section{Data Acquisition and Reduction} \label{sec:data}

Observations were obtained with the Atacama Large Millimeter/submillimeter Array (ALMA) pointed at the only site of active star formation in The Brick, the so-called ``maser core'' that was the focus of \citet{walker2021star}. These observations were taken in Cycle 5 as part of project 2019.1.00092.S (PI: Ginsburg). This project is a wideband, unbiased spectral line survey of The Brick, the goal of which is to identify unique molecular tracers for star formation processes (e.g., hot cores, shocks, outflows, dense gas) in extreme environments like the Galactic Center. The observations were for a single pointing with the 12-meter array, and seven correlator setups comprise the dataset across Bands 3, 4, and 6. The flux calibrator used was J1924-2914 and the phase calibrator was J1744-3116.

The frequency of the dasar transition of interest is in Band 3. At 107~GHz, the largest recoverable angular scale in our imaged data is 10$''$, the FWHM size of the synthesized beam is $1.48''$ by $1.03''$ with a position angle of $-74.03^\circ$, and the spectral resolution is 2.74~km~s$^{-1}$ (0.976 MHz at 107~GHz). Assuming a distance to The Brick of $\sim$8 kpc \citep{nogueras-lara2021distance}, $1''$ corresponds to $\sim$0.04 pc.

The observations were calibrated using the ALMA pipeline \citep{hunter2023alma}. We imaged the calibrated data into data cubes using CASA \citep{casateam2022}. The imaging script can be found on GitHub.\footnote{\url{https://github.com/abulatek/brick/tree/main/dasar/cleans.py}} 
We used the multiscale CLEAN algorithm, which helped recover the diffuse absorption features of the 107~GHz transition. To image the transition, we used the mask parameter \texttt{auto-multithresh} to automatically create a CLEAN mask based on where a(n absorption) signal appears on a per-channel basis \citep{kepley2020automultithresh}. We set the parameter \texttt{negativeThreshold} for \texttt{auto-multithresh}, which introduces an ``absorption mask'' that gets combined with the emission mask for each clean cycle. We used a \texttt{negativeThreshold} value of $-0.5\sigma$ to constrain the absorption mask in each channel. Our final cubes were created with natural weighting.

\subsection{Larger context}

\begin{figure*}
    \centering
    \includegraphics[height=1.12\textwidth]{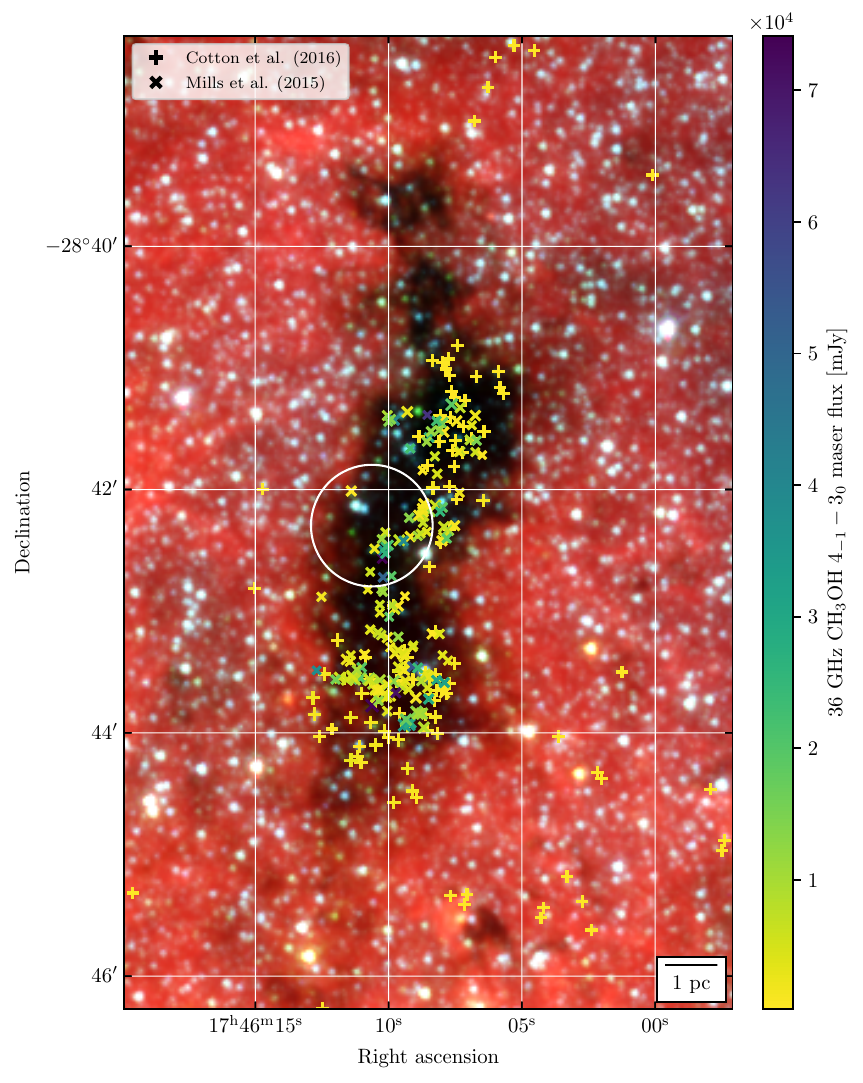}
    \caption{A red-green-blue (RGB) image of The Brick using the \textit{Spitzer} InfraRed Array Camera (IRAC) filters where red is IRAC 4 (8 $\mu$m), green is IRAC 2 (4.5 $\mu$m), and blue is IRAC 1 (3.6 $\mu$m). The data are from the \textit{Spitzer}/GLIMPSE surveys \protect\citep{benjamin2003glimpse}. The circle shows the extent of our pointing and has a radius of 30$''$ ($\sim$1 pc). The +'s and X's are locations of 36 GHz ($4_{-1} - 3_0\ E$) class I methanol masers from Table 2 in \protect\citet{cotton2016largescale} and Table 4 in \protect\citet{mills2015abundant} respectively. The size of the marker corresponds to the brightness temperature of the maser, and the color of the marker corresponds to the peak flux of the maser in mJy.}
    \label{fig:largercontext}
\end{figure*}

\begin{figure*}[h]
    \centering
        \includegraphics[width=0.44\textheight]{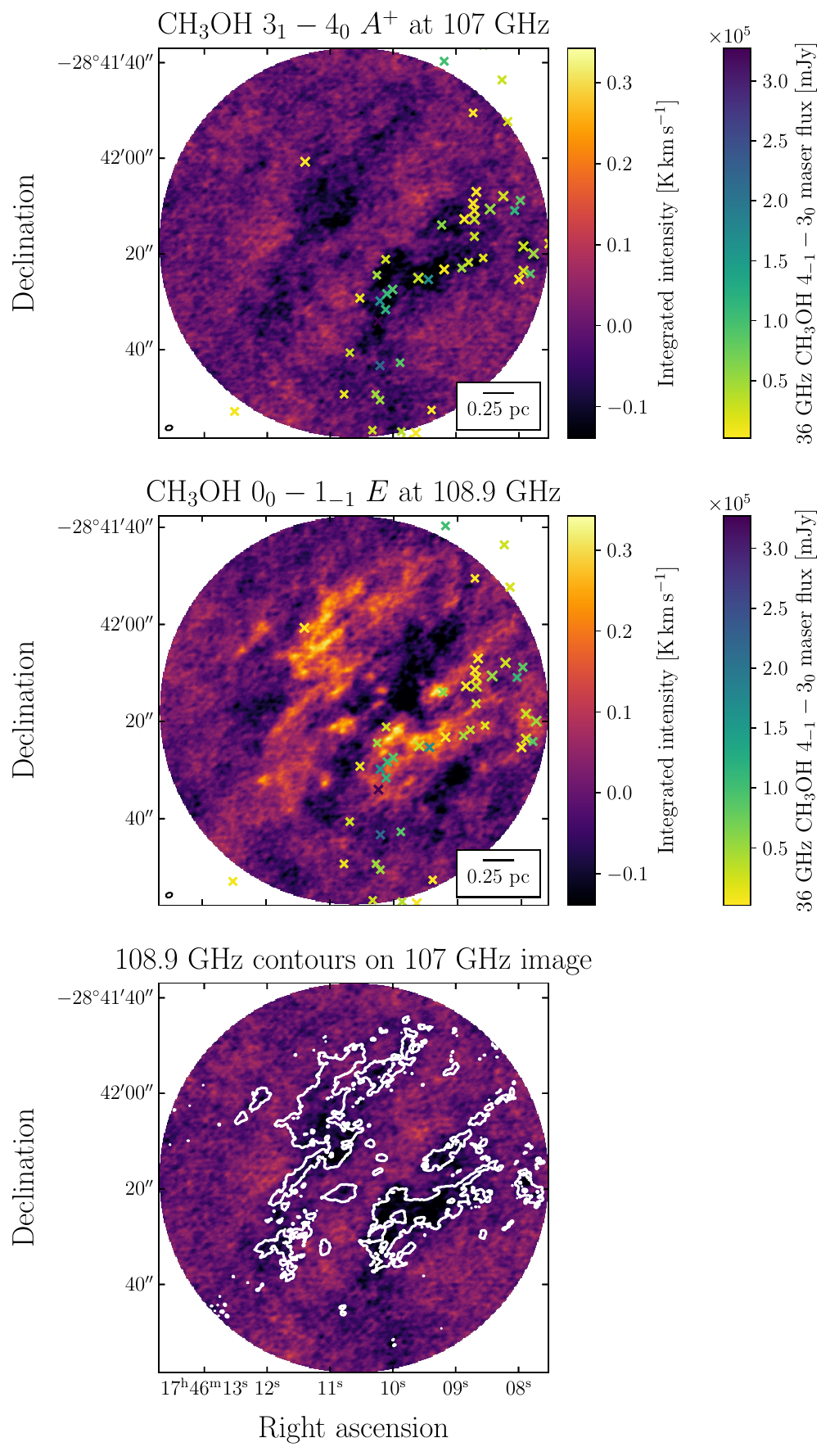}
    \caption{A comparison of the $3_{1} - 4_{0}$ $A^+$ 107~GHz methanol transition and the nearby $0_{0} - 1_{-1}$ $E$ 108.9~GHz methanol transition. The scale bars are 0.25 pc, or $\sim$$6.45''$ at these frequencies. \textit{Top:} an integrated intensity map of the 107~GHz methanol transition integrated from $-10$ to 25 km s$^{-1}$ in the LSRK velocity frame. The beam ($1.48''$ by $1.03''$) is shown in the lower left corner. For this panel and the middle panel, the \protect\citet{mills2015abundant} catalog of 36 GHz ($4_{-1} - 3_0\ E$) class I methanol masers is overlaid. The size of the marker corresponds to the brightness temperature of the maser, and the color of the marker corresponds to the peak flux of the maser in mJy. \textit{Middle:} an integrated intensity map of the 108.9~GHz methanol transition integrated over the same velocity range. The brightest patches are located where the 107~GHz integrated intensity map in the top panel shows the lowest level of emission. The beam ($1.37''$ by $0.997''$) is shown in the lower left corner. \textit{Bottom:} the same integrated intensity map of the 107~GHz methanol transition with contours over-plotted of the 108.9~GHz methanol transition. This panel shows the spatial correspondence of the 107~GHz absorption and the 108.9~GHz emission. These moment maps were made from cubes that were continuum-subtracted.
    }
    \label{fig:methanolcomparison}
\end{figure*}

Figure \ref{fig:largercontext} shows the FWHM of the 107~GHz primary beam centered on our pointing position within the larger context of The Brick. Our pointing is centered on a protostar with a visible outflow cavity, the edge of which is traced by, e.g., low-$J$ methyl cyanide, methanol, and transitions of other molecules \citep{walker2021star}. Compact continuum emission from the protostar itself is detected in the center of our pointing (it is unresolved and has a flux density of $\sim$3.5 mJy), but is not shown in the continuum-subtracted images in this work. Figure \ref{fig:methanolcomparison} shows the distributions of emission and/or absorption in several methanol transitions in the region surrounding the protostar. The 107~GHz transition shows larger-scale diffuse absorption outside of the central region. The $0_0 - 1_{-1}$ CH$_3$OH $E$-transition at 108.9~GHz shows emission at the same locations where the 107~GHz transition shows absorption. The position and size of the region shown is similar to that discussed by \citet{walker2021star} (e.g., Figure 5b in that work); their data showed multiple outflows from various continuum sources, but they are not the cause of the methanol absorption we see here.

In Figures \ref{fig:largercontext} and  \ref{fig:methanolcomparison}, we overlay two catalogs of 36 GHz class I \textit{E}-type methanol masers ($4_{-1} - 3_0$). The +'s are locations of masers in Table 2 of \citet{cotton2016largescale}, and the X's are locations of both confirmed and candidate masers in Table 4 of \citet{mills2015abundant}. The size of the marker corresponds to the brightness temperature of the maser, and the color of the marker corresponds to the peak flux of the maser in mJy.

\section{Results: Detection of a Dasar Transition} \label{sec:detection}

\subsection{Morphological comparison}

We detect the 107~GHz methanol transition in absorption. We can confirm that the transition is seen in absorption against the CMB using the following logic. There is no continuum emission detected in the interferometric data across most of the area where the methanol absorption is detected. This lack of continuum emission could be caused by resolved-out background continuum emission. However, we are able to rule out the influence of resolved-out continuum emission by looking at this region using a single-dish telescope. 
We looked at archival data of The Brick taken with the MUSTANG-2 instrument on the Green Bank Telescope \citep{ginsburg2020mustang} and saw significant large-scale continuum emission in this region. 
The mean brightness temperature in a $62.9''$ diameter aperture centered on our pointing, the same as ALMA's field of view at 90 GHz, is $\sim$0.03 $\pm$ 0.004~K, (an $\sim$8$\sigma$ detection). 
Further, by feathering Planck data with the MUSTANG-2 data, we can recover emission on even larger angular scales. Using the same measurement technique on the MUSTANG-2 image of The Brick combined with data from Planck, we measure a brightness temperature of $\sim$0.06 $\pm$ 0.004~K in our pointing.
These brightness temperatures are low compared to the CMB, but are very significantly detected. Therefore, the continuum must be dominated by the CMB.

The simplest way to qualitatively determine whether the 107~GHz transition is dasing is through morphological comparison with another methanol transition. This is illustrated in Figure \ref{fig:methanolcomparison}, where we compare the 107~GHz transition with the nearby 108.9~GHz $0_{0} - 1_{-1}$ $E$ transition of methanol. The top panel shows the absorption at 107~GHz, while the middle panel shows emission in the same region by the 108.9~GHz transition. The lower panel shows contours of the 108.9~GHz map over the 107~GHz map, which demonstrates the close morphological correspondence between the transitions.

While there are ``negative bowl'' features in the integrated intensity map of the 108.9~GHz methanol transition, they appear between (and are generated by) nearby areas of bright emission. There are no such bright emission features in the integrated intensity map of the 107~GHz transition, which lends credence to the interpretation that the darker features we see in that transition are true absorption. Additionally, the 108.9~GHz emission features are cospatial with the 107~GHz absorption features (illustrated in the bottom panel of Figure \ref{fig:methanolcomparison}), while the 108.9~GHz negative bowls are not cospatial with any features in the 107~GHz map. If there was correspondence between the 108.9~GHz negative bowls and a feature in the 107~GHz map, we could interpret the 107~GHz absorption features as filtering due to missing large angular scales. In the absence of that spatial correspondence, the most likely interpretation for the 107~GHz features is absorption. 

\subsection{Spectral extraction}

\begin{figure*}
    \centering
    \includegraphics[width=\textwidth]{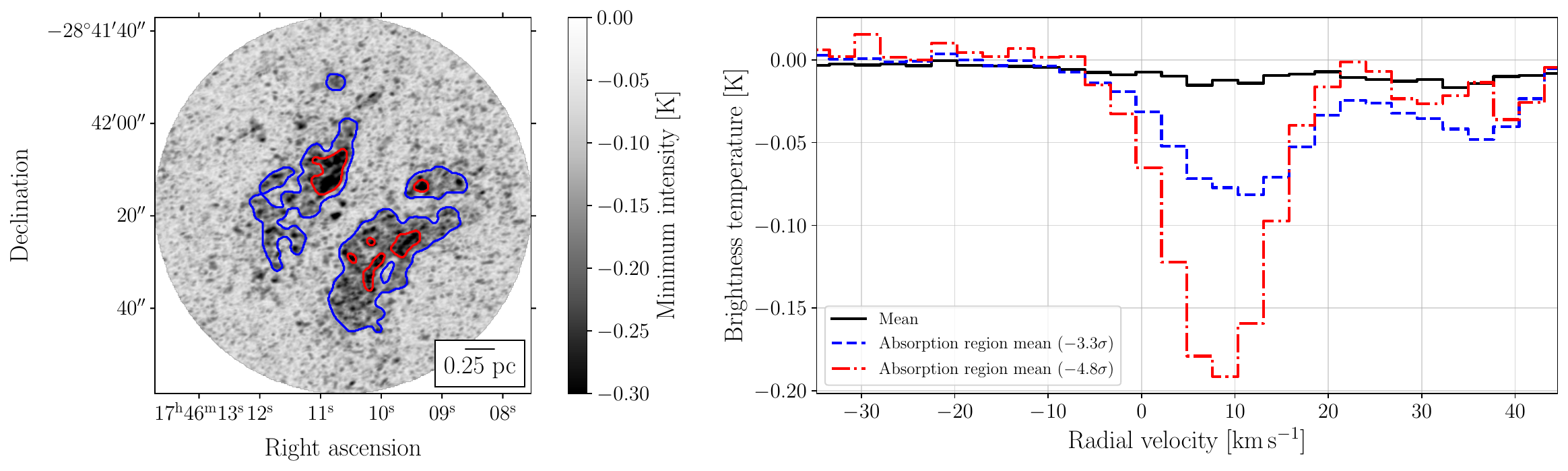}
    \caption{\textit{Left:} a minimum intensity map of the $3_{1} - 4_{0}\ A^+$ 107~GHz methanol transition. Two regions are shown to highlight the areas of absorption: one at $-3.3\sigma$ below a smoothed version of the minimum intensity map of the subcube (blue), and one at $-4.8\sigma$ (red). \textit{Right:} three spectra extracted from different parts of the pointing centered on the 107~GHz methanol transition. The solid black spectrum is the mean spectrum over the whole pointing; we see no net signal in that spectrum. The dashed blue and dot-dashed red spectra are mean spectra over the highlighted regions in the left panel, and they show deeper absorption than the mean spectrum over the whole pointing (about $-0.08$ K and $-0.2$ K for blue and red respectively). 
    The minimum intensity map was created from a continuum-subtracted cube.
    }
    \label{fig:spectralextraction}
\end{figure*}

To prove that the 107~GHz transition truly shows absorption and is not just darker in integrated intensity maps due to some interferometric artifact, we extracted spectra from both within the dasing region and outside of it. Figure \ref{fig:spectralextraction} shows this extraction. In the left panel, we show a minimum intensity map of the 107~GHz methanol transition. On that map, we show two regions covering the absorption. We derived these regions using the following procedure. 
\begin{enumerate}
    \item We performed a simple continuum-subtraction on a subcube centered on the 107~GHz transition by subtracting off the median of the subcube along the spectral axis ($-35$ to 45 km s$^{-1}$ in the LSRK velocity frame).
    \item We then calculated a minimum intensity map from the continuum-subtracted cube by taking the minimum along the spectral axis.
    \item We estimated the noise level $\sigma$ by taking the median absolute deviation (MAD) and scaling to the standard deviation.
    \item To make a mask that includes all absorption, we convolved the minimum intensity map with a 2D Gaussian kernel with a standard deviation of 5 pixels and made a smoothed mask of the map where the signal was below $-3.3\sigma$. We chose the most inclusive threshold that did not obviously bring in noise features.
    \item To remove isolated pixels that passed the S/N criterion, but are likely outliers, from the smoothed mask, we performed a series of binary dilations and erosions.\footnote{In binary dilation, for each pixel, if that pixel is False but has a neighbor that is True, assign it to be True. In binary erosion, for each pixel, if that pixel is True but has a neighbor that is False, assign it to be False.} We used 7 iterations of erosion followed by 7 iterations of dilation.
    \item To create a final mask, we used regions that are present in both the original smoothed mask and the eroded/dilated mask. This is the mask that is outlined in blue in the left panel of Figure \ref{fig:spectralextraction}. This union step removes pixels that were introduced into the mask by dilation that were not present in the original mask.
\end{enumerate}
We also repeated this process using a cutoff of $-4.8\sigma$ to capture the deepest parts of the absorption; in that process, we used only 2 iterations of erosion since the initial mask was more restrictive than the $-3.3\sigma$ mask. The $-4.8\sigma$ threshold was selected to create a mask that was as restrictive as possible but still included regions with enough pixels to average over. The region captured by the $-4.8\sigma$ mask is outlined in red in the left panel of Figure \ref{fig:spectralextraction}.

The right panel of Figure \ref{fig:spectralextraction} shows mean spectra extracted from the absorption regions (blue dashed line for $-3.3\sigma$ and red dot-dashed line for $-4.8\sigma$) as well as a mean spectrum over the entire pointing (solid line). We see real absorption at the level of $-0.08$ K averaged over the absorption region and at the level of $-0.2$ K in the darkest areas.

\section{Analysis: Excitation and Physical Conditions} \label{sec:radiative}

As discussed by \citet{Menten1991} and \citet{leurini2016physical}, in the absence of a strong infrared field, for $E$-type methanol, the $J_K$ rotational energy levels of the $K=-1$ ladder become overpopulated relative to levels in the neighboring $K=0$ and $K=-2$ ladders to which they can radiatively decay. This results in class I methanol maser action for $J > 4$, with the 36 GHz $4_{-1} - 3_0\ E$ transition being the lowest frequency example; see Table 1 of \citet{leurini2016physical} for a list of higher $J$ lines with $K=-1$ or $-2$. However, for transitions with  upper level $J\leq3$, the \textit{lower} level gets overpopulated, which results in dasar action in the $2_{0}-3_{-1}\ E$ transition observed by \citet{walmsley1988antiinversion} in dark clouds and also star-forming regions that contain  class I masers \cite[see Figure 1 of ][]{leurini2016physical}. An analogous situation arises for $A$-type methanol, where the $K = 0$ energy levels tend to possess the highest populations. Transitions with their upper energy level in the $K = 0$ ladder and their lower state in the $K = 1$ ladder, i.e., with levels $J \ge 7$ (Figure \ref{fig:energydiagram}), show class I maser action, with the 44 GHz $7_0-6_1\ A^+$ transition being the lowest frequency example (the cyan dashed line in the energy level diagram in Figure \ref{fig:energydiagram}). In contrast, transitions with levels with a lower $J$ are dasars. These include the 6.7 GHz $5_1-6_0\ A^+$ line and the 107~GHz $3_1-4_0\ A^+$ line discussed. Given the described mechanism, one expects dasar absorption also to occur in other lines (see Table \ref{tbl:lineinfo}; Figure \ref{fig:energydiagram} shows the transitions expected to dase, though there may be transitions at higher energy levels that might dase which we do not address in this paper). Moreover, one expects class I methanol maser emission from regions with enhanced absorption in the 107~GHz and other methanol dasar lines. 

\begin{figure*}
    \centering
    \includegraphics[width=\textwidth]{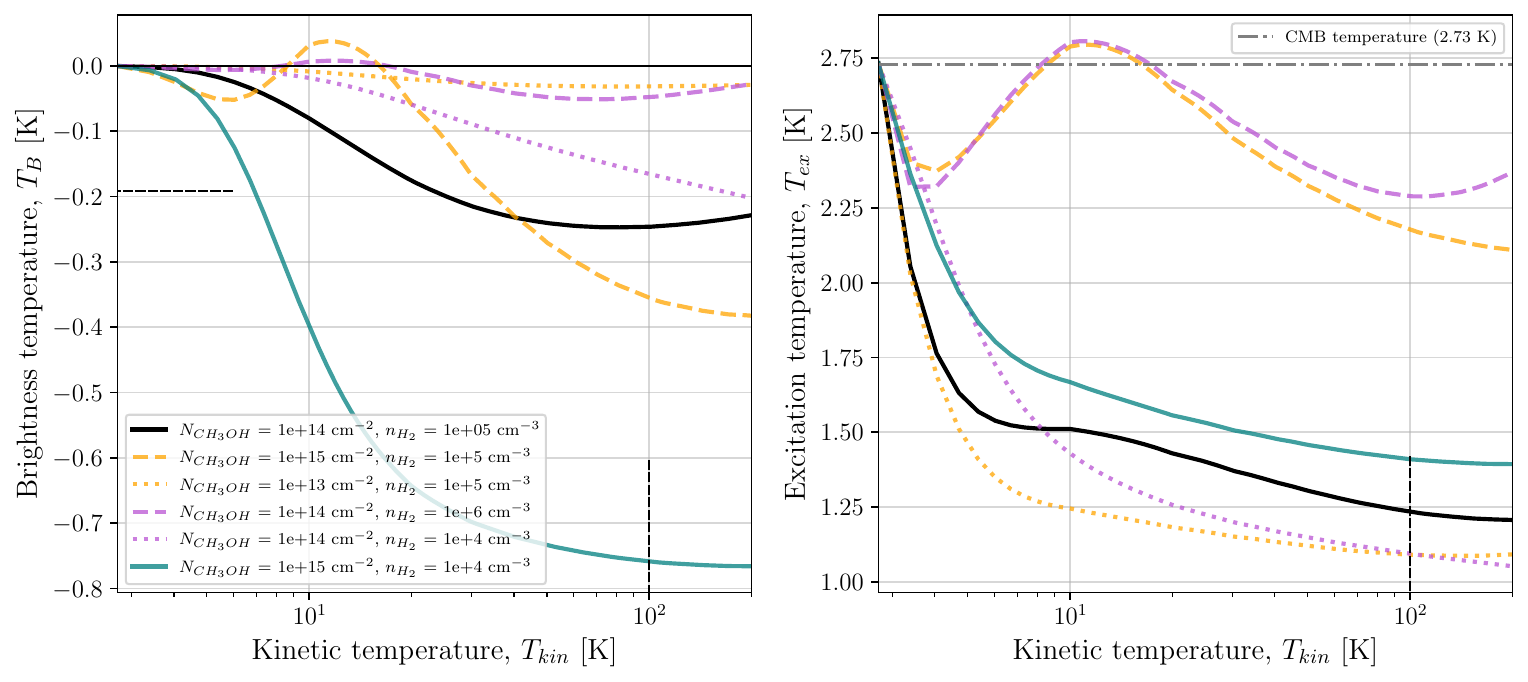}
    \caption{The behavior of the $3_{1} - 4_{0}\ A^+$ 107~GHz methanol transition (measured by brightness temperature $T_B$, \textit{left}, and excitation temperature $T_{ex}$, \textit{right}) as a function of kinetic temperature $T_{kin}$ in K at a variety of column/volume density combinations. The legend applies to both panels. The horizontal dashed line at $T_B \approx -0.19$ K is the minimum of the red dot-dashed spectrum in Figure \ref{fig:spectralextraction} and the vertical dashed lines at $T_{kin} = 100$ K show the kinetic temperature at which the modeling was performed.}
    \label{fig:temperature}
\end{figure*}

\subsection{Radiative transfer simulations}
To determine the underlying physical conditions under which we would see dasing, we performed radiative transfer simulations using a Python wrapper of RADEX\footnote{\url{https://home.strw.leidenuniv.nl/~moldata/radex.html}} \citep{vanderTak2007computer} called pyRADEX.\footnote{\url{https://pypi.org/project/pyradex/}} 
pyRADEX uses the Sobolev or large velocity gradient (LVG) approximation \citep[derived in][]{elitzur1992masers} for the escape probability geometry as a default, which we adopted.
We simulated the behavior of the 107~GHz methanol transition for a range of physical parameters (namely, number density and column density). We surveyed densities between $n_{H_2} = 10^3$ and $10^7$ cm$^{-3}$ and column densities between $N_{CH_3 OH} = 10^{13}$ and $10^{16}$ cm$^{-2}$.
All radiative transfer modeling was performed with a kinetic temperature of $T_{kin} = 100$ K \citep{johnston2014dynamics,ginsburg2016dense} and a uniform background temperature of $T_{CMB,0} = 2.7315$ K. 

We primarily used brightness temperature $T_B$ to look for dasing (a negative brightness temperature indicates dasing), though we also modeled excitation temperature and optical depth for each combination of density/column density.

We briefly explored the behavior of the 107~GHz transition at different kinetic temperatures in Figure \ref{fig:temperature} (2.73 K to 200 K, calculated with a variety of fixed $N_{CH_3 OH}$ and $n_{H_2}$ values). 
The lower limit of temperatures was chosen because 2.73 K is the coldest we would expect material in a $z = 0$ molecular cloud to be due to the CMB backlight; the upper limit of temperatures was selected because there is evidence of high temperatures in The Brick \citep{johnston2014dynamics}.

For the fixed column density $N_{CH_3 OH} = 10^{14}$ cm$^{-2}$ and volume density $n_{H_2} = 10^5$ cm$^{-3}$ (black solid line), the molecule exhibits dasing behavior at all surveyed kinetic temperatures, $2.73 < T_{kin} < 200$ K ($T_B$ is below 0 for all $T_{kin}$). For column and volume densities one order of magnitude below those fixed values ($N_{CH_3 OH} = 10^{13}$ cm$^{-2}$ with $n_{H_2} = 10^5$ cm$^{-3}$ and $N_{CH_3 OH} = 10^{14}$ cm$^{-2}$ with $n_{H_2} = 10^4$ cm$^{-3}$, dotted lines), we see similar dasing behavior. At higher column density and lower volume density ($N_{CH_3 OH} = 10^{15}$ cm$^{-2}$ and $n_{H_2} = 10^4$ cm$^{-3}$; teal solid line), we see even stronger dasing. However, for column and volume densities one order of magnitude above the fixed values ($N_{CH_3 OH} = 10^{15}$ cm$^{-2}$ with $n_{H_2} = 10^5$ cm$^{-3}$ and $N_{CH_3 OH} = 10^{14}$ cm$^{-2}$ with $n_{H_2} = 10^6$ cm$^{-3}$, dashed lines), dasing stops between $\sim$10 and $\sim$20 K before resuming at higher temperatures. 
Because dasing is enabled for this transition at nearly all temperatures we surveyed, we conclude that the effect does not strongly depend on temperature.

\begin{figure*}[h]
    \centering
    \includegraphics[width=\textwidth]{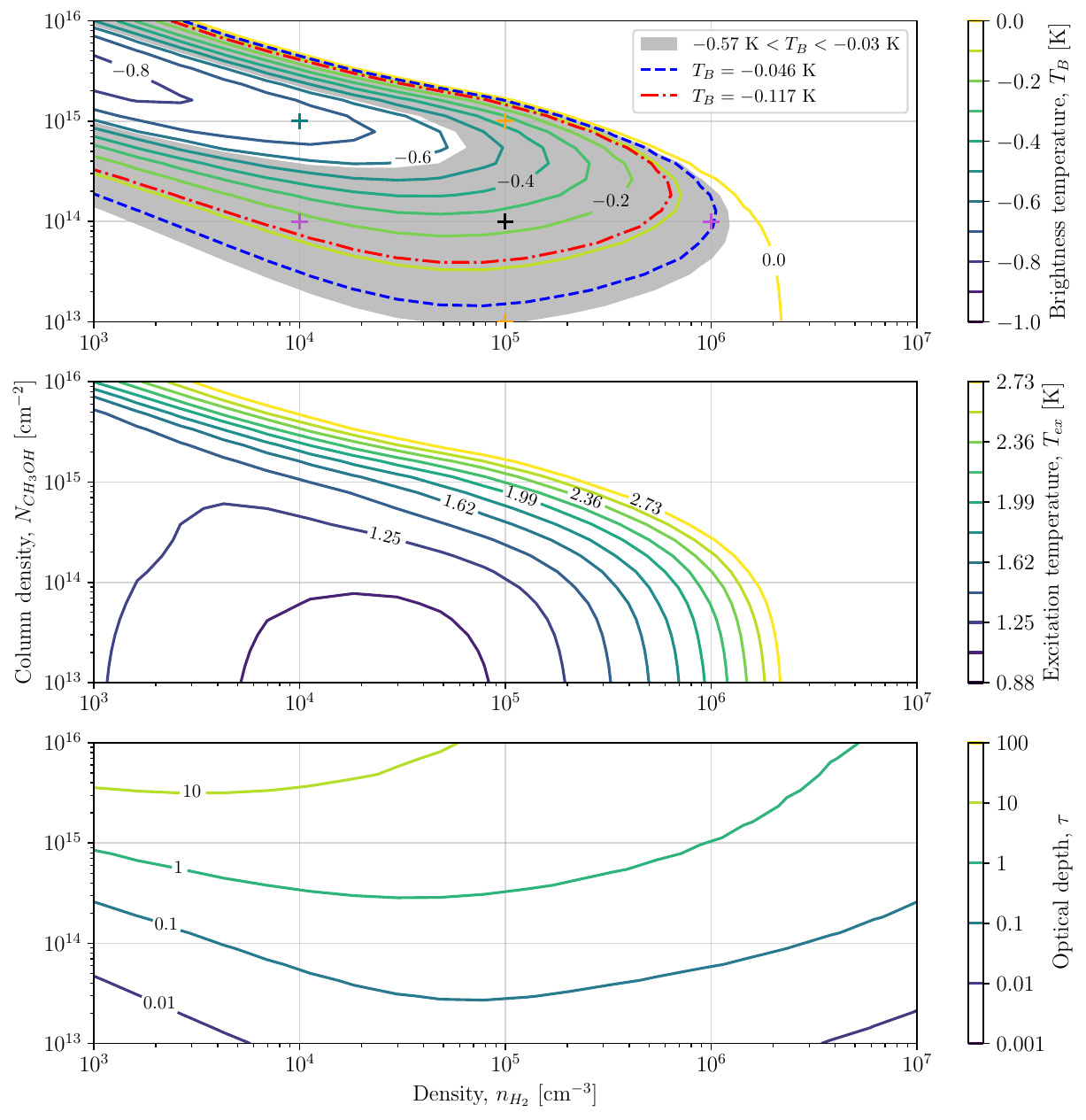}
    \caption{An exploration of the the behavior of the $3_{1} - 4_{0}\ A^+$ 107~GHz methanol transition as a function of two physical parameters: column density $N_{CH_3 OH}$ in cm$^{-2}$, and density $n_{H_2}$ in cm$^{-3}$. Each panel of this figure shows a different property modeled on a grid of column density and density values. All radiative transfer modeling was performed with a kinetic temperature of $T_{kin} = 100$ K. \textit{Top:} the modeled brightness temperature $T_B$ of the 107~GHz methanol transition in K. The shaded region shows where we observe dasing, i.e., when $-0.57$ K $< T_B < -0.03$ K (between the minimum value of the integrated intensity map and $-1\sigma$ in that map). The blue dashed and red dot-dashed lines show the minimum values from the corresponding spectra extracted from the absorption regions from Figure \ref{fig:spectralextraction} ($-0.046$ K and $-0.117$ K respectively). The +'s represent the parameter space we sample in Figure \ref{fig:temperature}. \textit{Middle:} the modeled excitation temperature $T_{ex}$ of the transition in K. 
    \textit{Bottom:} the modeled optical depth $\tau$ of the transition.}
    \label{fig:pyradexresults}
\end{figure*}

Figure \ref{fig:pyradexresults} shows the range of physical parameters over which the 107~GHz transition is expected to dase: densities below 10$^6$ cm$^{-3}$ and column densities between 10$^{13}$ and 10$^{16}$ cm$^{-2}$. The deepest absorption occurs at high column densities and low volume densities. An edge-on spiral galaxy is an example of an environment that could create this combination of observables.

\subsection{Potential methanol masers and dasars}

Other methanol transitions can also dase. Radiative transfer modeling predicts that two other methanol transitions should undergo dasing at typical physical parameters that enable dasing at 107~GHz ($N_{CH_3 OH} = 10^5$ cm$^{-2}$, $n_{H_2} = 10^{14}$ cm$^{-3}$, and $T_{kin} = 100$ K). We highlight these transitions in Figure \ref{fig:energydiagram}. For those physical conditions, the $2_1 - 3_0$ transition at $\sim$157~GHz has a predicted $T_B = -0.168$ K, and the $1_1 - 2_0$ transition at $\sim$206~GHz has a predicted $T_B = -0.0164$ K. Both are weaker than the predicted 107~GHz brightness temperature for the same physical conditions ($T_B = -0.246$ K). These are frequencies that are observable with ALMA, though they are not in-band for our survey of The Brick.

Since we examined the distribution of 36 GHz methanol masers across our pointing in Figure \ref{fig:methanolcomparison} and found some overlap between the \citet{mills2015abundant} sample and the dasar absorption we observed, we attempted to model the parameter range over which the 36 GHz transition mases. Across the same surveyed volume and column density range as in Figure \ref{fig:pyradexresults}, we find negative excitation temperatures and optical depths across the board, except for at $N_{CH_3 OH} > 5\times10^{15}$ cm$^{-2}$, indicating masing at nearly all surveyed column/volume density combinations. For the fixed column density $N_{CH_3 OH} = 10^{14}$ cm$^{-2}$ and volume density $n_{H_2} = 10^5$ cm$^{-3}$, the 36 GHz transition mases at all temperatures above $T_K = 5$ K.

Though the modeling shows that the 36 GHz transition mases at nearly all surveyed densities, we only see weak 36 GHz masers associated with one area within the dasar absorption region (see Figure \ref{fig:largercontext}). We are not sure why the masers are weaker here than outside of our pointing, nor why they appear to be only on one side of the protostar, but we note that there is significant variation in the strength of the masers surveyed by \citet{cotton2016largescale} and \citet{mills2015abundant}, so it is possible there is additional maser action in the region.

We also modeled the behavior of the 108.9~GHz transition shown in Figure \ref{fig:methanolcomparison}. It does not mase across the parameter range surveyed.

\section{Discussion: Dasars as a Tool} \label{sec:applications} 

\subsection{Cosmological applications}

One of the observational powers of a dasar comes from its ability to be detected with a constant signal-to-noise as redshift increases. 
This is because the strength of a dasar depends on the temperature of the CMB, which increases with redshift:
\begin{equation} \label{eqn:tz}
    T_{CMB} = T_{CMB,0} (1 + z)
\end{equation}
Here, $T_{CMB,0}$ is the local CMB temperature \citep[assumed to be 2.73 K for the following modeling;][]{fixsen2009temperature} and $T_{CMB}$ is the CMB temperature at redshift $z$.

\begin{figure*}
    \centering
    \includegraphics[width=\textwidth]{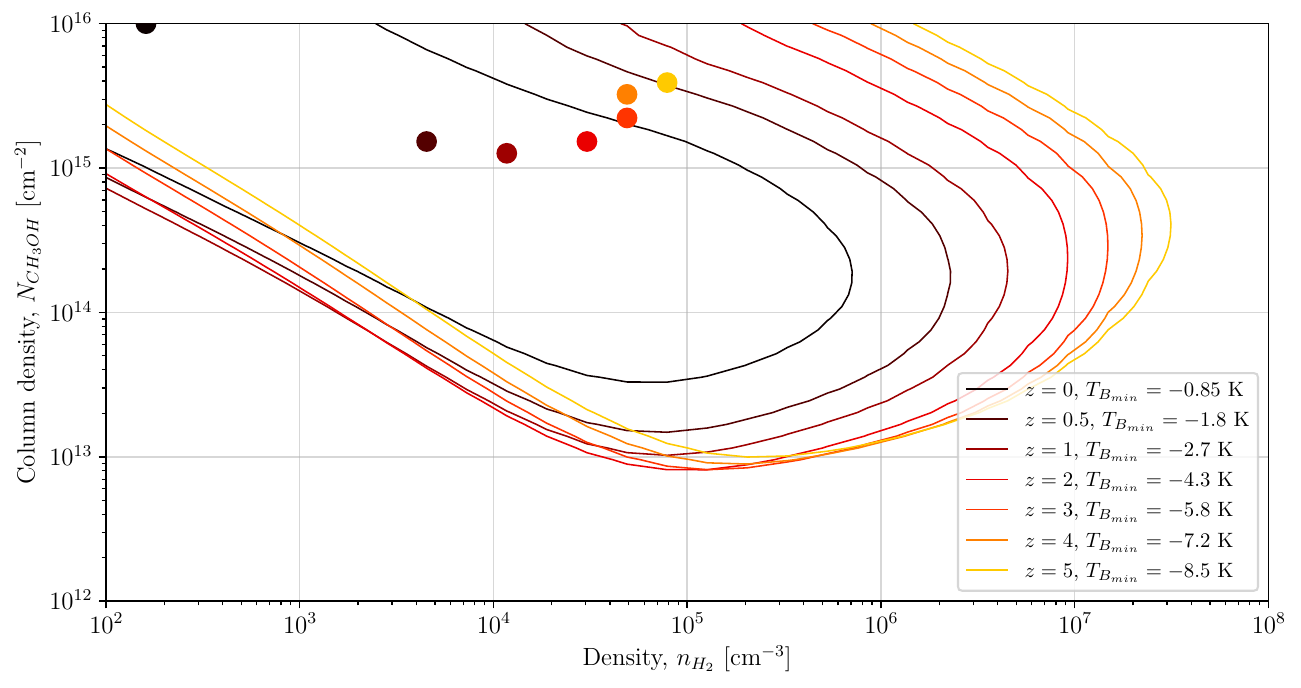}
    \caption{The effect of a varying CMB temperature (and thus redshift) on observable densities and column densities. All radiative transfer modeling was performed with a kinetic temperature of $T_{kin} = 100$ K. Absorption at the level of $T_B = -0.1$ K is shown using a solid line for each redshift ($z = 0$, 0.5, 1, 2, 3, 4, and 5). As redshift increases, the range of measurable physical parameters widens for a fixed level of absorption. 
    The circles mark the maximal absorption depth in the surveyed density/column density space, and the value of that absorption for each redshift is given in the legend.
    With the exception of $z=0$, for a deeper level of absorption, the maximal absorption depth moves towards higher densities and column densities as redshift increases.}
    \label{fig:redshift}
\end{figure*}

\begin{figure*}
    \centering
    \includegraphics[width=\textwidth]{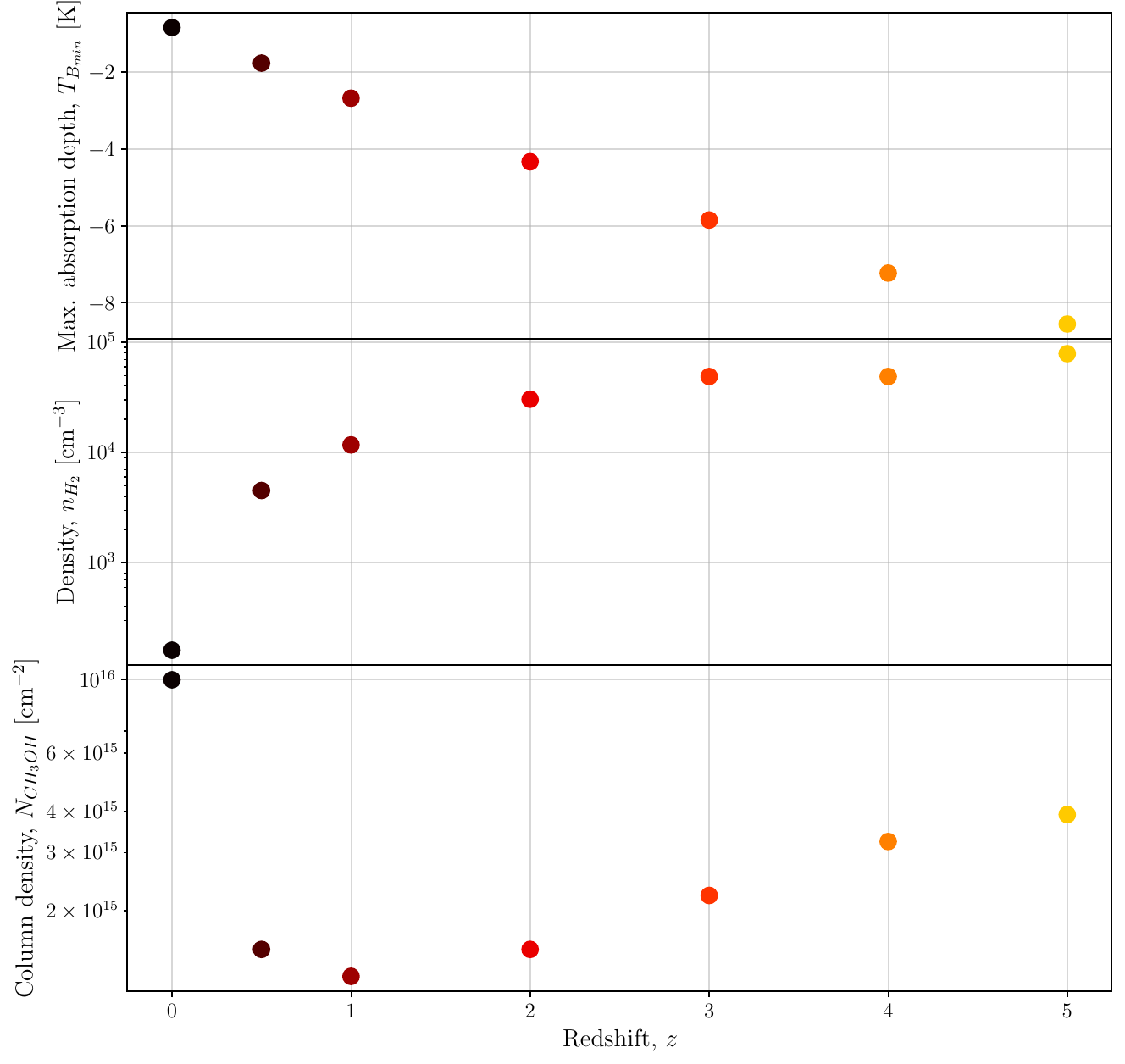}
    \caption{\textit{Top:} the evolution of maximal absorption depth $T_{B_{min}}$ with redshift $z$. \textit{Middle:} the evolution of the density at which the maximal absorption depth occurs as a function of redshift. \textit{Bottom:} the evolution of the column density at which the maximal absorption depth occurs as a function of redshift.}
    \label{fig:redshiftparams}
\end{figure*}

This changing CMB temperature means that as redshift increases, the physical parameter ranges that the dasar is tracing will also change. In Figure \ref{fig:redshift}, we demonstrate the effect of varying redshift on the observable threshold ($-0.1$~K, shown by solid contours) of the 107~GHz methanol dasar. At this threshold, as redshift (and therefore $T_{CMB}$) increases, the range of measurable physical parameters widens. For each redshift, we found the minimum brightness temperature, indicated by a circle. At this maximal level of absorption, the range of measurable physical parameters generally moves towards modestly higher densities and column densities as redshift increases. These trends are illustrated in Figure \ref{fig:redshiftparams}. As an exception, at $z=0$, the maximal absorption depth occurs at the upper limit of surveyed column densities. We also sampled higher column densities, but found that the models seemed to exhibit inconsistent behavior as a function of input parameters above $N_{CH_3 OH} \sim 5 \times 10^{16}$ cm$^{-2}$. We attribute this behavior to a failure of the numerical solver to converge.

\subsection{The 107~GHz transition is detectable in other galaxies}

The 107~GHz methanol transition can be seen as a dasar in other galaxies. \citet{humire2022methanol} showed absorption of the CMB by methanol at 107~GHz in two regions of the galaxy NGC 253 (R1 and R9, which are both located along the inner Lindblad resonance; the absorption spectra appear in Figure 2 in that paper). We inspected the non-continuum-subtracted cube at 107~GHz (priv. comm.) and while there exists extended continuum that overlaps spatially with the absorption feature, the absorption is deeper than the maximum from NGC 253. This is only possible if the absorption is against the CMB, or if the backlight is a continuum component larger than the largest angular scale observable with ALMA; we suggest the latter interpretation is unlikely. Thus, we can conclude that we see absorption against the CMB and not a continuum source from the Galaxy.

From the radiative transfer results (Figure \ref{fig:pyradexresults}), we can conclude that the 107 GHz dasar will select for galaxies that are dominated by gas with densities from $10^4$ to $10^6$ cm$^{-3}$. For denser gas, emission from the 107~GHz line will fill in the absorption from lower density gas. So, we can conclude that emission regions in NGC 253 must have $n_{H_2} > 10^6$ cm$^{-3}$ (or $N_{CH_3OH} \gg 10^{15}$ cm$^{-2}$).

Seeing the absorption occurring at the edges of the central molecular zone (CMZ) of NGC 253, where the column density is high due to the viewing geometry, seems to confirm our prediction from radiative transfer modeling that dasing should be stronger at higher column densities (Figure \ref{fig:pyradexresults}).

In order for this dasar to be a valuable tool, we must ensure that there are no nearby lines that could contaminate the absorption and make its measurement difficult or impossible. An emission line of HOCO$^+$ at $\sim$106.9 GHz is close in frequency to the 107~GHz methanol line. Depending on the %rotation curve of 
velocity structure and, in particular, the rotation curve of the galaxy, there is concern that this HOCO$^+$ transition could contaminate the signal from the 107~GHz methanol transition. If, from an inclined angle, we observe several velocities of material along the line of sight, and if two lines are close enough in rest frequency, they could overlap in their observed frequency. For any given spiral galaxy, we can determine the critical velocity difference between the center of the galaxy and its arms to check if there will be overlap between two lines:
\begin{equation}
    v_{max} \sin{i} \geq \Delta v_{CH_3 OH - contaminant}
\end{equation}
Here, $v_{max}$ is the maximum permissible velocity difference between the center of the galaxy and any exterior point, $i$ is the inclination of the galaxy, and $\Delta v_{CH_3 OH - contaminant}$ is the velocity difference between the target methanol line and the contaminant line. For this HOCO$^+$ potential contaminant line, $v_{max} \sin{i} \gtrapprox 280$ km s$^{-1}$. Assuming an inclination of $i \approx 75^\circ$ \citep{humire2022methanol}, $v_{max} \gtrapprox 290$ km s$^{-1}$. Because the rotation curve of NGC 253 in that region is characterized by velocities below 290 km s$^{-1}$, we find that the HOCO$^+$ line does not overlap with the 107~GHz methanol line. 

In addition to the absorption in the arm, we noticed the presence of emission at 107~GHz towards the center of NGC 253, which can be seen in the spectra for R3 through R7 in Figure 2 of \citet{humire2022methanol}. This emission might indicate that the dasar transition cannot be used to measure physical properties in highly-starbursting galaxies like NGC 253 that are area-dominated by very dense regions, as any absorption could be filled in by the emission from the very dense regions.

\subsection{Abundance measurements using dasars}

By combining the physical parameter constraints presented in Figure \ref{fig:pyradexresults} with dust continuum measurements, we can make a coarse estimate of the abundance of CH$_3$OH in The Brick. The abundance of a compound is defined by the ratio of its column density with respect to the column density of some other compound, usually H$_2$:
\begin{equation}
    X_{CH_3OH} \equiv \frac{N_{CH_3OH}}{N_{H_2}} 
\end{equation}
\citet{tang2021aztec} produced maps of $N_{H_2}$ using 1.1 mm dust continuum measurements of the Milky Way CMZ. At the location of our pointing, $N_{H_2} \approx 10^{23.3}$ cm$^{-2}$. Using a representative methanol column density from our modeling of $N_{CH_3 OH} = 10^{14}$ cm$^{-2}$, we estimate a methanol abundance of $X_{CH_3 OH} = 10^{-9.2}$ in this region of The Brick.

\subsection{Methanol dasar deep field survey}

As shown in \citet{darling2012formaldehyde} and \citet{darling2018formaldehyde}, a dasar transition can be a valuable tool for a redshift-unbiased deep field survey of regions of dense, star-forming gas in distant galaxies. Because methanol is a slightly asymmetric rotor like formaldehyde, multiple dasar transitions can be used in tandem as a densitometer to derive a unique density solution \citep{mangum2013densitometry}.

\begin{figure*}
    \centering
    \includegraphics[width=\textwidth]{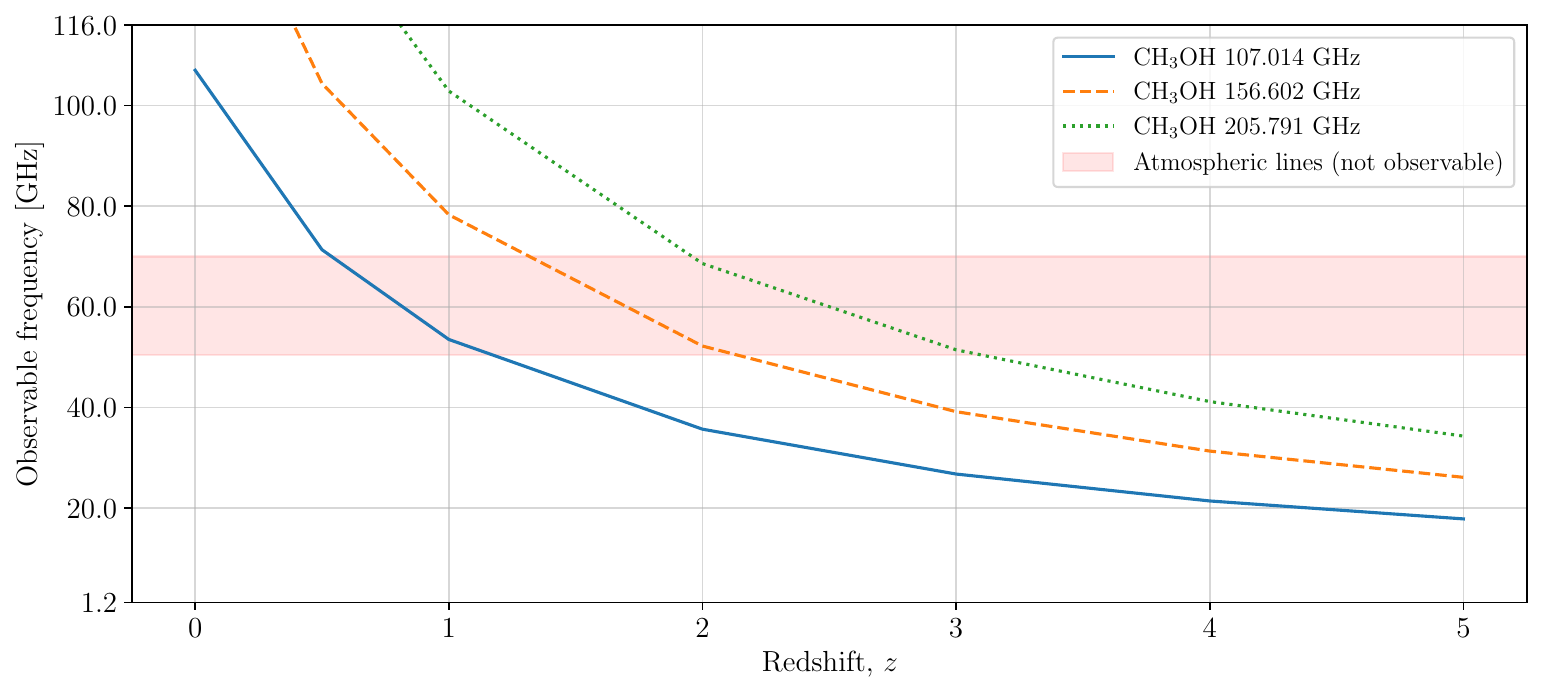}
    \caption{Methanol dasar transitions observable with the ngVLA as a function of redshift. The frequency range is limited to the target range for the ngVLA ($\sim$1.2 to 116 GHz). Other methanol dasars at higher frequencies may also be redshifted into the observable range. The horizontal shaded region shows the portion of the frequency range that is not covered by receivers (50.5 to 70 GHz) due to the presence of atmospheric absorption lines at those frequencies.}
    \label{fig:deepfield}
\end{figure*}

The 107~GHz methanol transition is particularly well-suited for a survey with the upcoming next-generation Very Large Array (ngVLA), as are the two other methanol transitions that are expected to dase at submillimeter wavelengths. Figure \ref{fig:deepfield} illustrates the observability of the $\sim$107, $\sim$157, and $\sim$206~GHz methanol transitions as a function of redshift. Above $z = 1$, all three lines are redshifted to within the frequency range of the ngVLA, which is estimated to be $\sim$1.2 to 116 GHz.

The following performance information is referenced in the ngVLA Memo 106.\footnote{\url{https://library.nrao.edu/public/memos/ngvla/NGVLA_106.pdf}; some metrics are also cited in tables at \url{https://ngvla.nrao.edu/page/performance}.}

This survey could make use of the entire frequency range offered by the ngVLA. The field of view of the ngVLA ranges from $\sim$25$'$ at 2.4 GHz to $0.628'$ at 93 GHz. Therefore, at higher frequencies (lower redshifts), multiple pointings would likely be needed to cover a typical galaxy. 

We can take NGC 253 as a typical starbursting galaxy. The regions that dase there are about 5$''$ by 5$''$. Assuming a distance to NGC 253 of 3.5 Mpc \citep{rekola2005distance}, they are about 100 pc on a side. Using the angular size-redshift relation,
this translates\footnote{\url{https://www.kempner.net/cosmic.php}} to $\sim$$0.012''$ at $z=1$, assuming a flat LCDM cosmology with $H_0 = 70$ km s$^{-1}$ Mpc$^{-1}$, $\Omega_m = 0.3$, and $\Omega_\Lambda = 0.7$. This resolution is achievable across the entire frequency range of the telescope, which means we can beam-match observations to the size of star-forming regions at $z=1$.

At this $0.012''$ resolution, the line sensitivity for one hour of exposure in a 10 km/s channel varies over the frequency range of the interferometer. It reaches a maximum at either end ($\sim$38.36 \textmu Jy/beam at 2.4 GHz, $\sim$39.00 \textmu Jy/beam at 93 GHz) and a minimum of $\sim$21.24 \textmu Jy/beam at 27 GHz.

The angular size-redshift relation \citep[see Figure 12 in][]{sahni2000case} ensures that if the ngVLA can resolve and beam-match dasing regions in a typical starbursting galaxy at $z=1$, it can resolve these regions at $z > 1$. 
The angular size of a 100 pc-size object decreases sharply until about $z=1$, reaching a minimum of just above $0.01''$, then stays nearly constant from $1 < z < 3$ before starting to increase again. This relation emerges due to the expansion of the universe. At earlier times, objects in the sky were closer to us than they are now. Due to the finite speed of light, the objects appear larger on the sky than they should based on their distance because the light we observe from those objects was emitted when the objects were much closer to us (and thus appeared larger on the sky).

\section{Conclusions} \label{sec:conclusions}

As part of a line survey of the Galactic Center molecular cloud The Brick, we detected absorption against the CMB by methanol, a phenomenon called a ``dasar.'' This absorption is seen in the $3_1 - 4_0\ A^+$ transition of methanol at $\sim$107~GHz. The absorption reaches a depth of $T_B \approx -0.2$ K in the deepest areas of the absorption region. We used pyRADEX to determine the physical conditions under which dasing occurs at this level, and determined that densities of H$_2$ between 10$^3$ and 10$^6$ cm$^{-3}$ and column densities of CH$_3$OH between 10$^{13}$ and 10$^{16}$ cm$^{-2}$ enable dasing.

We evaluated the potential of this dasar transition as a cosmological tool and showed that the physical parameter space probed by the transition favors higher densities and column densities at higher redshifts. We also discussed the detection of this dasar transition in the starburst galaxy NGC 253. This detection proves that this methanol line is detectable in other galaxies. The 107~GHz methanol dasar transition, as well as two other dasar transitions of methanol which are observable at radio frequencies at $z > 1$, are good candidates for a deep field survey with the next generation of radio interferometers, e.g., the ngVLA.

\begin{acknowledgments}
We thank the anonymous referees for their helpful feedback.  
This paper makes use of the following ALMA data: ADS/JAO.ALMA\#2019.1.00092.S. ALMA is a partnership of ESO (representing its member states), NSF (USA) and NINS (Japan), together with NRC (Canada), MOST and ASIAA (Taiwan), and KASI (Republic of Korea), in cooperation with the Republic of Chile. The Joint ALMA Observatory is operated by ESO, AUI/NRAO and NAOJ. 
The National Radio Astronomy Observatory is a facility of the National Science Foundation operated under cooperative agreement by Associated Universities, Inc.
AG acknowledges support from the NSF under grants AST 2008101, 2206511, and CAREER 2142300.
\software{CASA \citep{casateam2022}, 
\texttt{astropy} \citep{astropy:2013,astropy:2018,astropy:2022},
\texttt{spectral-cube} \citep{spectralcubezenodo},
\href{https://radio-beam.readthedocs.io/en/latest/}{\texttt{radio-beam}}
}
\facility{ALMA}
\end{acknowledgments}

\bibliography{refs}{}
\bibliographystyle{aasjournal}

% \allauthors
\end{document}